# EXCITONS WITH CHARGE TRANSFER IN $SnCl_2$-PHTHALOCYANINE FILMS

YA.I. VERTSIMAKHA, P.M. LUTSYK

Institute of Physics, Nat. Acad. Sci. of Ukraine *(46, Nauky Prosp., Kyiv 03680, Ukraine)*

The absorption, modulated photoreflectance, and photovoltage spectra of dichlorotin phthalocyanine ($SnCl_2Pc$) films have been measured. These films are thermally deposited in vacuum at different substrate temperatures. The energies of charge-transfer-states (CT-states) in $SnCl_2Pc$ films (1.35, 1.52, and 2.05 eV) and the diffusion length of Frenkel excitons (130±30 nm) have been determined. The photosensitivity of $SnCl_2Pc$ films is comparable to that of *n*-type perylene derivative (MPP) layers and by several times ($\approx$ 5) more than the photosensitivity of thermally deposited *n*-type $C_{60}$ films.

1. Introduction

Phthalocyanines (Pc) belong to the most stable and photosensitive organic semiconductors (OS) [1, 2] and, for a long time, are widely used for a creation of photoconverters as the -type component [3, 4]. At the same time, the number of photosensitive organic *n*-type semiconductors that are required for the development of *p* - *n* heterostructures and solar cells is limited. In contrast to the majority of OS (including tin Pc [5]), diclorotin phthalocyanine ($SnCl_2Pc$) is an *n*-type semiconductor [6, 7]. Photovoltaic properties of $SnCl_2Pc$ films haven't been investigated, to our knowledge.

Furthermore, the cubic nonlinear optical susceptibility {$\chi^{(3)}$} of $SnCl_2Pc$ films is 3 orders greater than those of other Pc films [8]. Probably, such a rising of $\chi^{(3)}$ is a result of the resonance coincidence of the laser excitation energy with that of one of the CT-states or the energy of double singlet-triplet transitions. However, now these energies haven't been determined.

The transitions of CT-states are forbidden by symmetry in Pc with planar molecular structure (zinc Pc, copper Pc). These transitions can be observed only in electroabsorption spectra [9, 10]. If the ionic radius of the central atom (Pb, Sn) or atomic group (VO, TiO) of a Pc molecule is large, then the central atom is outside of the plane of a Pc ring [7, 8]. These Pc molecules have a strongly reduced symmetry in comparison with that of a planar Pc molecule. In addition, at the growth of films, the non-planar Pc molecules can form a thermodynamically efficient structure (modification).

In this structure, the central atom of one molecule is located opposite to peripheral hydrogen atoms of the Pc ring of another molecule. In that case, the induced electrostatic interaction appears between these molecules. The energy of this interaction is about or more than the Van der Vaals interaction energy. These both reasons lead to the strong increase of the probability of CT-states transitions. As a consequence, CT-states transitions can be observed in absorption and photovoltage spectra [13]. Since the size of a $SnCl_2$ group is large, a $SnCl_2Pc$ molecule is considered to be non-planar.

Therefore, the aim of the present article was the investigation of the optical and photovoltaic properties of $SnCl_2Pc$ films and the estimation of a contribution of CT-states to electronic processes in these layers.

2. Experimental Procedure

$SnCl_2Pc$ films were prepared by thermal deposition in vacuum (2 x $10^{-4}$ Pa) from a powder. The $SnCl_2Pc$ powder was purified by vacuum sublimation with the use of a tantalum evaporator. The films were deposited on quartz substrates without and with a conductive transparent indium tin oxide (ITO) layer. The substrate temperature ($T_a$) was 300 and 410 K. The thickness of the films was controlled during the deposition by a change in the frequency of a quartz sensor. After the films preparation, the thickness was measured with an atomic force microscope (AFM). The thickness of obtained $SnCl_2Pc$ films was 180 and 800 nm.

The absorption spectra (AS) were measured by using a double-beam "Hitachi" spectrophotometer. The measurements of modulated photoreflectance were carried out by a technique described in [14, 15]. The laser excitation with 1.96-eV energy of quanta (*hv*) was used in these measurements.

The photovoltage was measured with the Bergman technique modified and improved by Akimov [16].

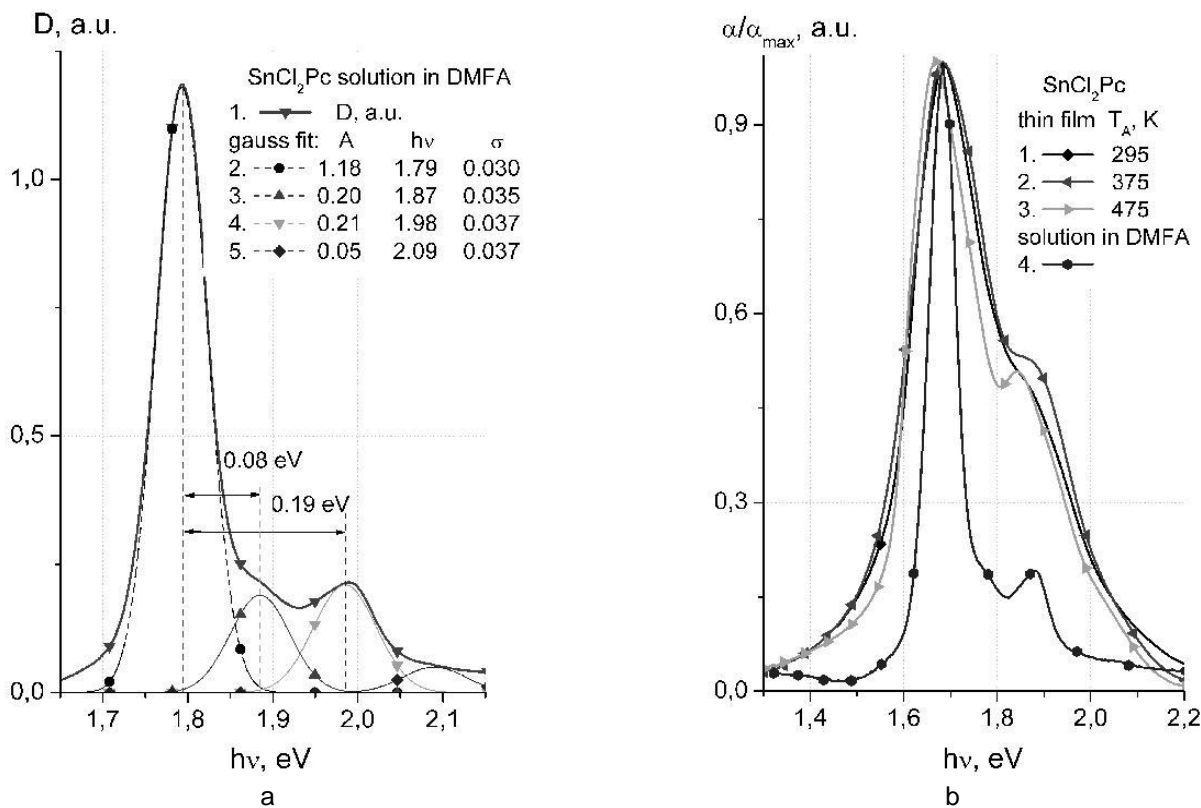

Fig. 1. *a* Spectral dependence of the optical density, *D*, of an SnCl$_2$Pc solution in DMFA (*1*) and the Gauss fit (*2-5*); *b* AS normalized on the maximum of SnCl$_2$Pc films with thickness 180 nm before (*1*) and after annealing at $T_a$ = 375 K (*2*), 475 K (*3*) and the SnCl$_2$Pc solution in DMFA (*4*) is red-shifted by 0.12 eV

## 3. Results

The AS of a researched SnCl$_2$Pc solution in dimethylformamide (DMFA) is practically coincide with the AS of other SnCl$_2$Pc solutions [17] and is well described by the electronic transition with the 1.79-eV energy and weak bands with intermolecular oscillations at 0.08 and 0.19 eV (Fig. 1,*a*). The AS of SnCl$_2$Pc films with a thickness of 180 nm are similar to that of the solution in DMFA, though the AS maximum of SnCl$_2$Pc films (1.67 eV) is red-shifted by 0.12 eV relative to the AS maximum of the solution. The half-widths of the absorption bands of SnCl$_2$Pc films are greater than that of an SnCl$_2$Pc solution (Fig. 1,*b*). The half-widths of the absorption bands of SnCl$_2$Pc films decrease at the annealing in air during one hour. This sharpening of bands increases at annealing temperature ($T_a$) rising in the range 330 475 K. The observed annealing effect on the AS of SnCl$_2$Pc films agrees with experimental data obtained in [18].

The AS of an SnCl$_2$Pc solution (red-shifted by 0.12 eV) and films are identical in the range of strong absorption (1.65 1.8 eV) only (Fig. 1,*b*). Absorption of SnCl$_2$Pc films is more than the red-shifted absorption of the solution in the ranges of weak absorption (1.2 1.6 and 1.9 2.2 eV). This indicates the additional absorption appearance in SnCl$_2$Pc films.

The spectra of the modulated photoreflectance $\Delta R$=$R(h\nu)$ were measured to estimate the bands energies of the additional absorption in SnCl$_2$Pc films. The peaks of $\Delta R$=$R(h\nu)$ at (1.34±0.02) and (1.55±0.02) eV are showed for SnCl$_2$Pc films annealed at $T_a$ = 375 and 475 K in the range 1.15 - 1.65 eV (Fig. 2,*a*). The magnitude of peaks at (1.34±0.02) eV increases at a rise of $T_a$. The magnitude of the minimum at (1.55±0.02) eV decreases with increase in $T_a$.

To confirm the appearance of additional absorption bands, the AS of SnCl$_2$Pc films of 800 nm in thickness were measured. The optical densities, $D(h\nu)$, of annealed SnCl$_2$Pc films increase in the range 1.30 1.38 eV and decrease in the range 1.4 1.6 eV.

As $T_A$ increases, the differences of $D(h\nu)$ for SnCl$_2$Pc films, $\Delta D(h\nu)$, have a maximum at (1.34±0.02) eV

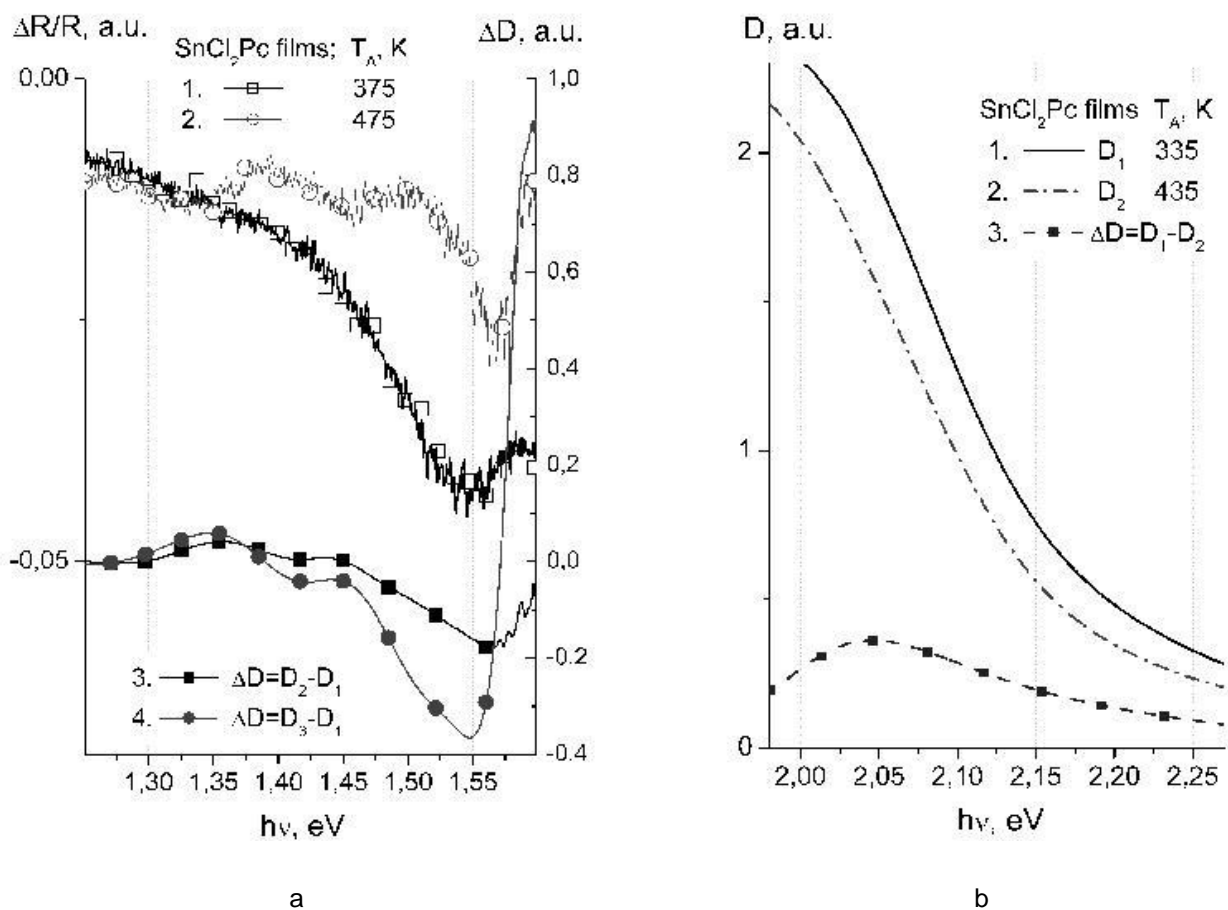

Fig. 2. *a* spectra of modulated photoreflectance, $\Delta R/R$, for $SnCl_2Pc$ films with a thickness of 180 nm annealed at $Ta$ = 375 K (*1*) and 475 K (*2*) and the difference of optical densities $\Delta D(h\nu)$, at $Ta$ increasing from 335 K to 435 K (*3*) and to 475 K (*4*) for $SnCl_2Pc$ films with a thickness of 800 nm; *b* spectral dependence of the optical density, $D$, of $SnCl_2Pc$ films with a thickness of 800 nm annealed at $Ta$ = 335 K (*1*), 435 K (*2*) and its difference, $\Delta D(h\nu)$, at $Ta$ increasing from 335 K to 435 K (*3*)

(Fig. 2,*a*) and minima at $(1.55\pm0.02)$ eV (Fig. 2,*a*), $(2.05\pm0.02)$ eV (Fig. 2,*b*). The intensity of the maximum increases at $T_a$ increasing and, on the contrary, the intensity of the minima decreases.

The photovoltage (in the visible and near IR ranges) of *n*-type structures of $SnCl_2Pc$ and MPP were close and about 5 times more than the photosensitivity of thermally deposited $C_{60}$ layers [19].

The photovoltage spectra of ITO/$SnCl_2Pc$ structures at the illumination of a free surface, $\varphi_s$, and ITO contact, $\varphi_c$, are close and correlate with the absorption coefficient of ITO/$SnCl_2Pc$ structures, $\alpha(h\nu)$, in the range of weak absorption ($\alpha < (3-6)\times10^4$ cm$^{-1}$) (Fig. 3). $\varphi_c < \varphi_s$ and $\varphi_c(h\nu)$ anticorrelate with $\alpha(h\nu)$ in the range of strong absorption ($\alpha > 6 \times 10^4$ cm$^{-1}$) (Fig. 3). At $\alpha < 3 \times 10^4$ cm$^{-1}$, $\varphi(\alpha)$ is linear at the illumination of different sides of ITO/$SnCl_2Pc$ structures (Fig. 4). At $\alpha > 4 \times 10^4$ cm$^{-1}$ $\varphi_s(\alpha)$ tend to the saturation, while $\varphi_c(\alpha)$ decreases as $\alpha$ increases (Fig. 4).

4. Discussion

The red shift of the AS of $SnCl_2Pc$ films relative to the AS of an $SnCl_2Pc$ solution is caused by the interaction of molecules in a solid state. The statistical crystallite disorder and the appearance of the additional absorption in $SnCl_2Pc$ films can result in a broadening of the AS in the films (as compared to those in an $SnCl_2Pc$ solution) (Fig. 1,*b*). The appearance of the additional absorption in $SnCl_2Pc$ films compared to a red-shifted $SnCl_2Pc$ solution is associated with the formation either impurity states during the deposition or CT-states. Pc molecules are stable and don't decompose in the used temperature range. Moreover,

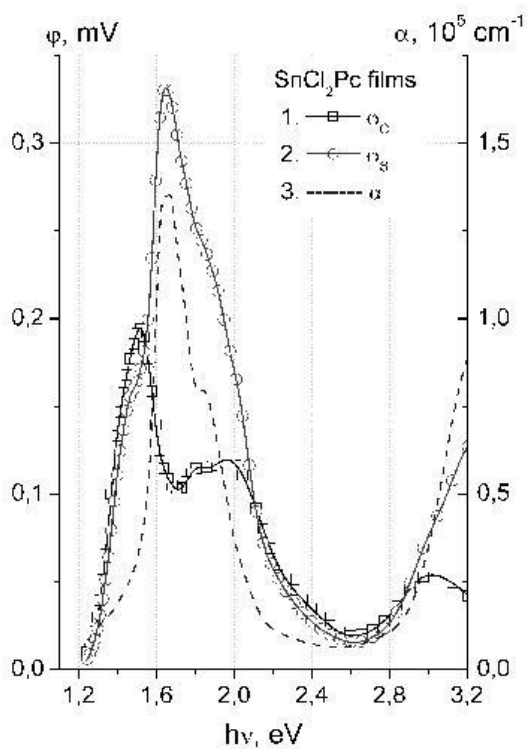

Fig. 3. Spectral dependence of photovoltage at illumination of ITO contact, $\varphi c$, (*1*) and free surface, $\varphi s$, (*2*) and absorption coefficient, $\alpha$, (*3*) of ITO/SnCl$_2$Pc structures at $T_s$ = 415 K

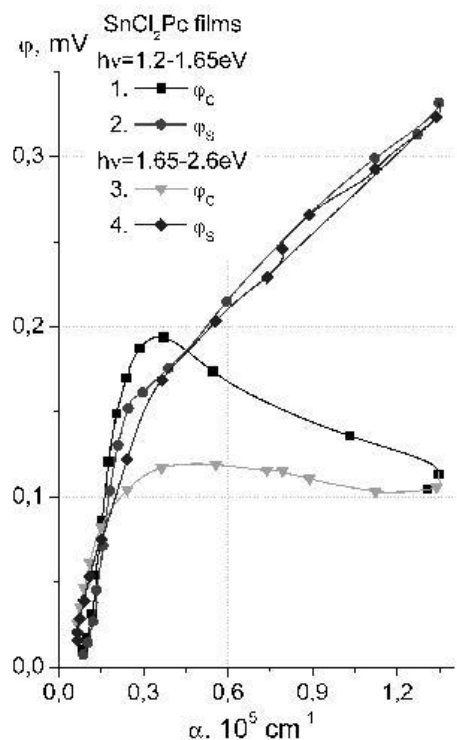

Fig. 4. Dependence of the photovoltage on the absorption coefficient, $\varphi(\alpha)$ at the illumination of the ITO contact (*1,3*) and the free surface (*2,4*) of ITO/SnCl$_2$Pc structures in the ranges 1.20-1.65 eV (*1,2*) and 1.65-2.60 eV (*3,4*)

the additional purifying from other uncontrollable impurities occurs during the vacuum deposition. Therefore, the formation of impurities is improbable during the deposition process. CT-states appear as a result of the hydrogen-like interaction of central atoms of one molecule with peripheral H-C atoms of Pc rings of another molecule.

The CT-states formation strongly depends on the crystalline modification of films. Two CT-states were seen for each crystalline modification in Pc films with non-planar molecular structure: PbPc, VOPc, and TiOPc. The energy of some CT-states was red-shifted, and the energy of other CT-states was blue-shifted relative to the energy of singlet excitons in Pc films [9]. The structure of SnCl$_2$Pc films have two crystalline modifications (triclinic and monoclinic) [20, 21]. Consequently, two CT-states with red-shifted energy and two CT-states with blue-shifted energy relative to the singlet exciton energy should be observed in SnCl$_2$Pc films. The AS of SnCl$_2$Pc films consist of both red-shifted absorption bands of molecules with maxima at 1.65, 1.73, 1.84, and 1.92 eV and the additional absorption bands associated with the formation of CT-states in SnCl$_2$Pc films with maxima at 1.35, 1.55, and 2.05 eV. These peaks were found out, by using the $\Delta R = R(h\nu)$ (Fig. 3) and $\Delta D(h\nu)$ (Figs. 4 and 5) spectra. It is probable that the intensity of the absorption bands of CT-states with the energy $h\nu > 1.65$ eV is small. Since the bands number and their maximum energy for AS of SnCl$_2$Pc films are known, it is possible to estimate the intensity of these bands. To this end, we made the AS fitting for the films by using Gauss components. The fit results are presented in the Table.

The comparison of CT-states absorption bands intensities (A) and areas (Θ) in SnCl$_2$Pc films shows (the Table) that: the CT-states absorption bands intensities with maxima at 1.35 eV decrease and, on the contrary, the intensities with maxima at 1.52 and 2.05 eV increase as $T_a$ rises. This confirms that the bands with maxima at 1.35 eV are the result of the CT-states absorption of one modification, and the bands with maxima at 1.52 and 2.05 eV are associated with other modification.

The general contribution of the CT-states absorption bands in SnCl$_2$Pc films weakly decreases

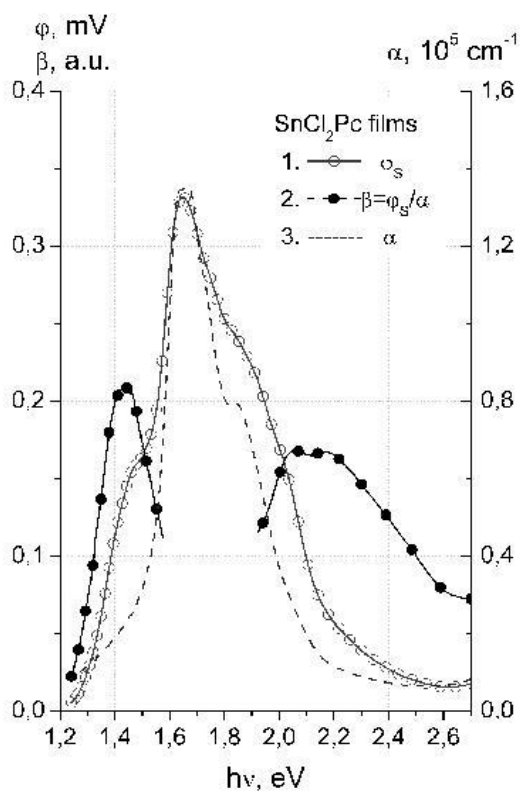
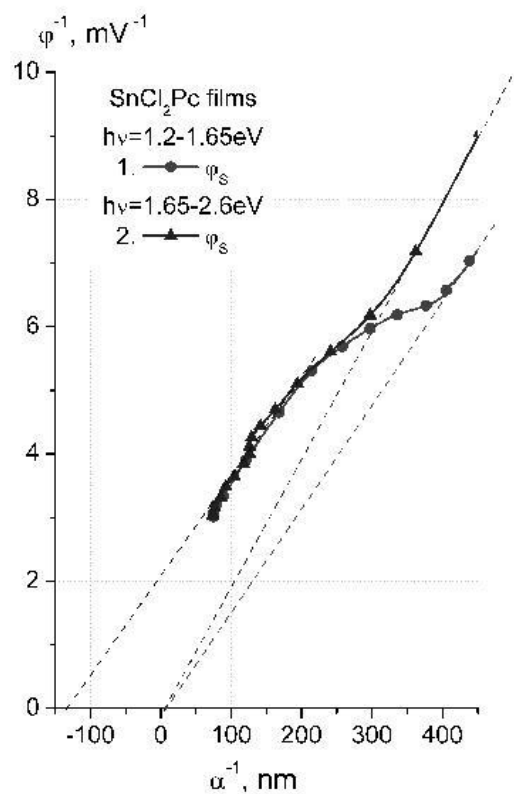

Fig. 5. Spectral dependence of photovoltage, φ, (*1*) absorption coefficient, *α*, (*2*), and quantum efficiency, β, (*3*) of ITO/SnCl$_2$Pc structures

Fig. 6. Dependence of $\varphi^{-1}(\alpha^{-1})$ in the ranges 1.20-1.65 eV (*1*) and 1.65-2.60 eV (*2*) at the illumination of the free surface of ITO/SnCl$_2$Pc structures

from 19 to 15% as $T_a$ increases. This can be associated with the fact that the probability of transitions with the participation of CT-states in the high-temperature crystalline modification is less than that in the low-temperature modification.

We compared the data on the investigated films and PbPc, VOPc, TiOPc films [13]. The contributions of CT-states in AS of SnCl$_2$Pc films and the monoclinic modification of this Pc are showed to be close (15-16%), but less than the contributions of CT-states in AS of the triclinic modification of VOPc and TiOPc films (21-29%). The last facts can be associated with the formation of the hydrogen-like interaction between O and H atoms of neighboring molecules for VOPc and TiOPc [13].

The correlation of φ($h\nu$) with α($h\nu$) and $\varphi_s \approx \varphi_c$ for SnCl$_2$Pc films in the range of weak absorption (α < (3 - 6) × 10$^4$ cm$^{-1}$) testifies to the equal bend of bands

Parameters of the Gauss fitting of the AS for SnCl$_2$Pc films annealed at different $Ta$

| | Number of a band | 1 | 2 | 3 | 4 | 5 | 6 | 7 |
|---|---|---|---|---|---|---|---|---|
| $Ta$ = 295 K | $h\nu_i$, eV | 1.35 | 1.52 | 1.653 | 1.73 | 1.84 | 1.922 | 2.05 |
| | $A_i$, a.u. | 0.06 | 0.185 | 0.83 | 0.80 | 0.41 | 0.33 | 0.14 |
| | $\Theta_i = A_i \cdot \sigma_i$, a.u. | 0.005 | 0.014 | 0.043 | 0.042 | 0.022 | 0.021 | 0.011 |
| | $\Theta_i/\Sigma\Theta_i$, % | 3.2 | 8.7 | 27.1 | 26.7 | 13.9 | 13.5 | 6.9 |
| $Ta$ = 375 K | $h\nu_i$, eV | 1.35 | 1.52 | 1.65 | 1.73 | 1.84 | 1.924 | 2.05 |
| | $A_i$, a.u. | 0.066 | 0.21 | 0.95 | 0.78 | 0.48 | 0.37 | 0.135 |
| | $\Theta_i = A_i \cdot \sigma_i$, a.u. | 0.006 | 0.014 | 0.048 | 0.04 | 0.025 | 0.022 | 0.01 |
| | $\Theta_i/\Sigma\Theta_i$, % | 3.4 | 8.3 | 29.5 | 24.2 | 15.5 | 13.3 | 5.8 |
| $Ta$ = 475 K | $h\nu_i$, eV | 1.35 | 1.52 | 1.65 | 1.731 | 1.84 | 1.925 | 2.05 |
| | $A_i$, a.u. | 0.09 | 0.19 | 1.38 | 0.75 | 0.64 | 0.37 | 0.13 |
| | $\Theta_i = A_i \cdot \sigma_i$, a.u. | 0.007 | 0.012 | 0.065 | 0.037 | 0.032 | 0.02 | 0.009 |
| | $\Theta_i/\Sigma\Theta_i$, % | 4.0 | 6.5 | 35.7 | 20.2 | 17.6 | 11.2 | 4.8 |

at the free surface and the ITO contact of ITO/SnCl$_2$Pc structures (Figs. 3 and 4).

The tendency to the saturation of φ$_s$(α) as α increases and the anticorrelation of φ$_c$(hv) with α(hv) (Figs. 6 and 7) in the range of strong absorption (α > 6 × 10$^4$ cm$^{-1}$) of SnCl$_2$Pc films evidence for one- and two-level recombination, respectively [22,23]. The direct one-level recombination of charge carriers occurs in the absence of trapping by impurity centers. As for the two-level recombination, it is a result of both the trapping of a non-equilibrium carrier by an impurity center and the subsequent trapping of a carrier of opposite sign by this charged center.

The direct one-level recombination of charge carriers is seen to occur at the illumination of the free surface of the investigated SnCl$_2$Pc films (Figs. 3 and 4) and is caused by the presence of charge carrier recombination centers on the surface. The anticorrelation of φ$_c$(hv) with α(hv) (Figs. 3 and 4) testifies to the two-level recombination of charge carriers by trapping centers. The trapping centers are formed on the interface of SnCl$_2$Pc and ITO layers. The small difference between φ(α) (Fig. 4) can be caused by the different contributions of CT-states in different spectral ranges.

The obtained data don't allow determining the nature of recombination and trapping centers. However, the recombination centers are considered to appear by the adsorption of active molecules of air (for example, oxygen) on the free surface. The trapping centers can be formed due to the diffusion of atoms from an ITO-electrode during the formation of SnCl$_2$Pc films.

According to [22-24],

$$\varphi = A \beta\alpha/(1+\alpha L), \qquad (1)$$

where A is the parameter independent of hv and is determined by the bend of bands on the illuminated side, β is the quantum efficiency of the photogeneration of charge carriers, and L is the diffusion length of excitons or CT-states. Therefore, in the weak absorption range (αL < 1), the quantum efficiency β(hv) ~ φ(hv)/α(hv).

The partial separation of charge carriers occurs under the excitation of CT-states. Therefore, the creation of free charge carriers requires a less energy, than that for the ionization of Frenkel excitons. Therefore, β should be more at the excitation of CT-states, than at the excitation of Frenkel excitons, if the energy difference between a CT-state level and the conduction band is small. Consequently, β(hv) should have maxima at energies, where the absorption is dominated by CT-states.

β(hv) of SnCl$_2$Pc films at $T_s$ = 415 K has maxima near 1.5 and 2.0 eV (Fig. 5, curve *3*). The energies of the maxima practically coincide with the absorption maxima of two CT-states (Fig. 2).

The band of CT-states at 1.35 eV isn't shown in β(hv), because the probability of the formation of free charge carriers is very small for the CT-state at 1.35 eV. This can be caused by a larger difference of the energies of this CT-state and the conduction band than that between the energies of the CT-states near 1.5 and 2.0 eV and the conduction band. Consequently, the efficiency of the photogeneration of charge carriers for the CT-states at 1.35 eV is less than that for excitons or another excitation of CT-states.

We see that relation (1) can be written as φ$^{-1}$(hv) ~ α$^{-1}$(hv)+ L. For Frenkel excitons, L can be determined from the dependence of φ$^{-1}$(α$^{-1}$) in the range of strong absorption (αL > 1). Because the Frenkel excitons give the basic contribution to the AS and photovoltage in this range, we get L = (130 ± 30) nm for SnCl$_2$Pc films. The estimated L for SnCl$_2$Pc films is more than that for photosensitive zinc Pc [25] or MPP [26] films, but less than L for pentacene [27] and hexathiopentacene [26] films. Unfortunately, it is impossible to estimate the diffusion length of CT-states in this way. The contribution of CT-states to α(hv) and φ(hv) dominates in the range of weak absorption (αL < 1), where φ$^{-1}$(hv) doesn't depend practically on L.

5. Conclusions

The weak bands at 1.35, 1.52, and 2.05 eV have been found in the absorption and modulated photoreflectance spectra of SnCl$_2$Pc films. These bands aren't associated with the absorption of molecules and can be caused by the formation of charge transfer states in SnCl$_2$Pc films. The formation of these states is a result of the hydrogen-like interaction of Cl atoms of one molecule with peripheral H-C atoms of phthalocyanine rings of another molecule. The contributions of CT-states in the absorption spectra of SnCl$_2$Pc films and the monoclinic modification of PbPc, VOPc, and TiOPc (15-16%) are close, but less than that of the triclinic modification of VOPc and TiOPc (21-29%).

The efficiency of the photogeneration of charge carriers at the excitation of charge transfer states at 1.52 and 2.05 eV is more than the efficiency of the photogeneration of charge carriers at the excitation of located singlet excitons. The diffusion length of excitons is (130 ± 30) nm in SnCl$_2$Pc films.

The photosensitivity of SnCl$_2$Pc films is comparable with that of *n*-type perylene derivative (MPP) layers widely used for the preparation of organic solar cells and is by ≈ 5 times more than that of thermally deposited *n*-type C$_{60}$ films.

The authors are grateful to J. Misiewicz and A. Podhorodecki from Wroclaw Technical University for their help in the measurements and interpretation of the photoreflectance spectra of the studied films.

ЕКСИТОНИ З ПЕРЕНОСОМ ЗАРЯДУ В ПЛІВКАХ SnCl$_2$-ФТАЛОЦІАНІНУ

*Я.І. Верцімаха, П.М. Луцик*

Р е з ю м е

Проведено вимірювання спектрів поглинання, фотомодульова-ного відбиття та фото-ерс плівок SnCl$_2$-фталоціаніну, отрима-них вакуумною сублімацією на підкладки з різною темпера-турою. Визначено енергії екситонів з переносом заряду (1,35; 1,52 і 2,05 еВ) і довжину дифузії екситонів Френкеля ((130±30) нм) в них. Фоточутливість плівок SnCl$_2$-фталоціаніну за вели-чиною порівнянна з фоточутливістю шарів *n*-типу перилено-вих похідних (MPP) і приблизно у 5 разів більша за фоточут-ливість термічно напилених шарів C$_{60}$ *n*-типу.